\documentclass[runningheads]{llncs}
\usepackage[T1]{fontenc}
\usepackage[inline]{enumitem}

\usepackage{graphicx}
\usepackage{xcolor}
\usepackage{xurl}
\usepackage{listings}
\usepackage{subcaption}
\usepackage{hyperref}
\usepackage[misc,geometry]{ifsym}

\usepackage{color}

\definecolor{codegreen}{rgb}{0,0.6,0}
\definecolor{codegray}{rgb}{0.5,0.5,0.5}
\definecolor{codeblue}{rgb}{0,0,1}
\definecolor{backcolour}{rgb}{1,1,1}

\lstdefinelanguage{ProVerif}{
  classoffset = 1,
  morekeywords = {among, axiom, channel, choice, clauses, const, def, diff, do, elimtrue, else, equation, equivalence, event, expand, fail, for, forall, foreach, free, fun, get, if, implementation, in, inj-event, insert, lemma, let, letfun, letproba, new, noninterf, noselect, not, nounif, or, otherwise, out, param, phase, pred, proba, process, proof, public vars, putbegin, query, reduc, restriction, secret, select, set, suchthat, sync, table, then, type, weaksecret, inj, yield},
  keywordstyle = \color{codeblue},
  morecomment = [s]{(*}{*)},
  commentstyle = \color{codegreen},
}
\lstdefinestyle{proverifstyle}{
    backgroundcolor=\color{backcolour},   
    commentstyle=\color{codegreen},
    keywordstyle=\color{codeblue},
    numberstyle=\tiny\color{codegray},
    stringstyle=\color{black},
    basicstyle=\ttfamily\footnotesize,
    breakatwhitespace=true,         
    breaklines=true,                 
    captionpos=b,                    
    keepspaces=true,                 
    numbers=left,                    
    numbersep=5pt,                  
    showspaces=false,                
    showstringspaces=false,
    showtabs=false,                  
    tabsize=2
}
\begin{document}
\renewcommand{\thelstlisting}{\arabic{lstlisting}}

\title{Misbinding Raw Public Keys to Identities in TLS}
%
%

\author{Mariam Moustafa \Letter \orcidID{0009-0003-3046-8675} \and
    Mohit Sethi\orcidID{0000-0002-9730-1955} \and
    Tuomas Aura\orcidID{0000-0003-1648-8875}}
\authorrunning{M. Moustafa et al.}

%
\institute{{Aalto University, Espoo, Finland} \\
\email{\{firstname.lastname\}@aalto.fi}}
\maketitle              

\begin{center}
\textbf{Accepted Manuscript} \\
This version of the article has been accepted for publication in NordSec 2024. The final version is available online at SpringerLink.
\end{center}

\begin{abstract}
  The adoption of security protocols such as Transport Layer Security (TLS) has significantly improved the state of traffic encryption and integrity protection on the Internet. Despite rigorous analysis, vulnerabilities continue to emerge, sometimes due to fundamental flaws in the protocol specification. This paper examines the security of TLS when using Raw Public Key (RPK) authentication. This mode has not been as extensively studied as X.509 certificates and Pre-Shared Keys (PSK). We develop a formal model of TLS RPK using applied pi calculus and the ProVerif verification tool, revealing that the RPK mode is susceptible to identity misbinding attacks. Our contributions include formal models of TLS RPK with several mechanisms for binding the endpoint identity to its public key, verification results, practical scenarios demonstrating the misbinding attack, and recommendations for mitigating such vulnerabilities. These findings highlight the need for improved security measures in TLS RPK.\footnote{This is the accepted manuscript version. The final published version is available at \url{https://doi.org/10.1007/978-3-031-79007-2_4}}
\end{abstract}


\section{Introduction}
\label{intro}

Development of standard security protocols, among which Transport Layer Security (TLS)~\cite{rfc8446} is the most prominent one, has significantly increased the usage of traffic encryption and integrity protection on the Internet~\cite{zscaler2023encrypted}. The protocols have undergone rigorous security analysis~\cite{bhargavan17}, ensuring their robustness and reliability. Furthermore, numerous open-source implementations of TLS and other protocols are readily available, facilitating their widespread adoption. Consequently, many products~\cite{sensity_tls,eimsig_tls}, services~\cite{surfshark_vpn_protocols,filezilla_sftp}, and standards~\cite{3gpp33501,RSP-01} now readily incorporate TLS and other security protocols as fundamental building blocks.

Despite the rigorous analysis these protocols undergo during their specification and the extensive scrutiny of their open-source implementations, new attacks are continuously discovered. Many of these attacks originate from weaknesses in the implementation of the protocols, rather than flaws in the protocols themselves~\cite{levillain2021implementation}. Nonetheless, attacks are also discovered from fundamental flaws in the protocol specification itself. For example, version 1.3 of TLS was specified at the Internet Engineering Task Force (IETF) after several years of deliberation and analysis. Yet, shortly after publication of the standard, an identity misbinding attack was discovered~\cite{drucker2021selfie}. 

TLS supports three independent modes of authentication: 
\begin{enumerate*}[label=(\roman*)] 
  \item X.509 certificates
  \item Pre-Shared Keys (PSK), and 
  \item Raw Public Keys (RPK)
\end{enumerate*}. 
While the security properties of TLS authentication with X.509 certificates and PSKs have been extensively analyzed in literature~\cite{bhargavan17,cremers17,drucker2021selfie}, a similar thorough analysis of the RPK authentication mode in TLS is missing. Therefore, in this paper, we study the security properties of TLS when authentication is based on raw asymmetric key pairs. We develop a formal protocol model of TLS RPK using applied pi calculus and the ProVerif verification tool~\cite{blanchet2018proverif}. Our formal verification of TLS RPK security properties shows that it is susceptible to identity misbinding attacks. Based on our findings, we provide practical recommendations on how the attacks can be avoided in practice. Thus, the contributions of this paper are three-fold:
\begin{itemize}
	\item Formal model of TLS RPK with different methods of binding the server and client identities to their public keys.
	\item Verification results that reveal server and client identity misbinding attacks, and example scenarios where the attacks can occur in practice.
	\item Discussion of the potential solutions available to the TLS RPK standards process and implementors.
\end{itemize}

The rest of the paper is organized as follows. Section~\ref{background} provides background information on the TLS RPK mode. Section~\ref{model} introduces our formal model of TLS RPK. Sections~\ref{misbinding} and~\ref{mutual} detail the formal verification results, including an analysis of the identified misbinding attacks. Section~\ref{sec:implementation} illustrates the ease of implementing these misbinding attacks in practice. Section~\ref{solution} discusses potential solutions to prevent the identified attacks. Finally, Section~\ref{conclusion} concludes the paper.


\section{TLS with Raw Public Keys}
\label{background}

This section provides background information on TLS and its raw public key mode, the security of which will be analyzed in the following sections.  

\subsection{The TLS Protocol}

Transport Layer Security (TLS) is one of the most long-standing security technologies on the Internet, providing a critical foundation for secure communications. TLS version 1.0~\cite{rfc2246} was standardized in 1999. For authentication, TLS adopted the web public key infrastructure (PKI)~\cite{rfc5280}, which is based on the X.509 standard~\cite{ituT_X509}. The certificates or certificate chains bind the endpoint public keys to their identities, such as host names. In typical web browsing, only the server has a certificate, but in other applications, the client can also have a certificate for mutual public-key authentication.

Over time, the protocol has evolved to improve its security and broaden its applications. As part of this evolution, new authentication methods have been defined. For instance, the Pre-Shared Key (PSK) authentication mode was introduced to avoid the computational expense of asymmetric operations and simplify key management in closed environments~\cite{rfc4279}.

As vulnerabilities were discovered in earlier versions, TLS underwent significant revisions. TLS version 1.2~\cite{rfc5246} focused on updating the cryptographic functions. Later, a \textit{raw public key} (RPK) mode was incorporated into TLS 1.2~\cite{rfc7250}, offering a lightweight alternative to certificate-based authentication. TLS 1.3~\cite{rfc8446} was a major update towards a more robust and formally verified protocol. It also incorporated three authentication methods: certificates, PSK, and RPK. Additionally, EAP authentication~\cite{rfc9190} was defined as an extension.

\subsection{Raw Public Keys in TLS}

The motivation for introducing the TLS raw public key (TLS RPK) was to reduce message size and processing cost. The \textit{Certificate} message in the RPK mode contains only the \textit{SubjectPublicKeyInfo} object~\cite{rfc7250}, which comprises an algorithm identifier and the raw public key. Compared to the full X.509~\cite{ituT_X509} certificates and certificate chains, this significantly reduces the amount of data transmitted and simplifies the processing at the endpoint that receives the key, which usually is the TLS client. Only a minimalist ASN.1 parser is needed instead of the complex and error-prone code for certificate validation~\cite{6956560,georgiev2012most,luo2023complexity}. The reduced code footprint can be particularly beneficial in resource-constrained IoT devices. The trade-off is that the endpoints must establish the authenticity of the raw public key in some alternative way without being able to rely on X.509 certificates. Thus, the RPK mode forgoes both the complexity and the benefits of the public-key infrastructure (PKI).

\begin{figure}[t]
    \begin{subfigure}[t]{0.50\textwidth}
        \centering
        \vspace{2pt}
        \includegraphics[scale=0.69]{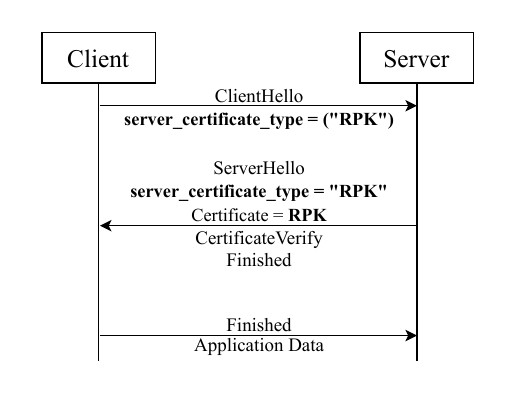}%
        \vspace{22pt}
        \caption{Server authentication}
        \label{fig:tls1.3-server-rpk}
    \end{subfigure}%
    \begin{subfigure}[t]{0.50\textwidth}
        \centering
        \vspace{0pt}
        \includegraphics[scale=0.69]{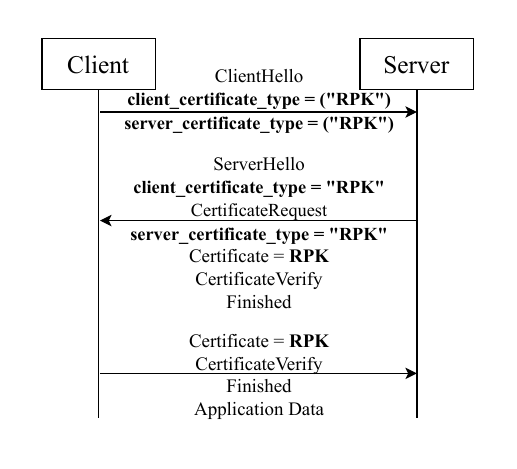}
        \caption{Mutual authentication}
        \label{fig:tls1.3-mutual-rpk}
    \end{subfigure}
    \caption{TLS 1.3 authentication with RPK}
    \label{fig:tls1.3-rpk}
\end{figure}

Figure~\ref{fig:tls1.3-server-rpk} illustrates TLS server authentication with RPK when the client is anonymous or authenticated separately inside TLS. The client sends the \lstinline{server_certificate_type} extension in the \textit{ClientHello} message. The extension informs the server which certificate types the client is willing to process~\cite{rfc7250}. In this case, the client asks for the \textit{RawPublicKey} type. The server sends back the \lstinline{server_certificate_type} extension in the \textit{ServerHello} message, indicating that it agrees to use the RPK mode. The server then sends its raw public key in the \textit{Certificate} message. While the raw keys are not certificates in the usual sense, the TLS specification treats them as one certificate type to unify the message structures in the different authentication modes.

TLS RPK also supports mutual authentication and has a corresponding \lstinline{client_certificate_type} extension. Figure~\ref{fig:tls1.3-mutual-rpk} shows a mutual RPK authentication where the client and server authenticate each other with raw public keys. The client first sends the \lstinline{client_certificate_type} extension in the \textit{ClientHello} message. This extension informs the server which certificate types the client can send. In the figure, the client only supports the \textit{RawPublicKey} type. The server agrees to the RPK mode for client authentication by responding with a \textit{CertificateRequest} for the same type. The client then sends a \textit{Certificate} message that contains its RPK.

\subsection{Authentication of the Raw Public Key}
\label{sec:peer-rpk}

In public-key authentication, there must be some way of binding the public key to the identity of its owner. Since the RPK mode does not use a PKI, there must be another solution for this binding. The TLS RPK specification~\cite{rfc7250} suggests several methods for a client or a server to authenticate the other endpoint's raw public key. 

\paragraph{Authentication with pre-configured keys} One possible solution is that the client and server have pre-configured lists of trusted public keys, which are configured by an out-of-band mechanism. While the standard does not provide further guidance, it is reasonable to assume that when there are multiple clients and servers, each of them has a local file or database indexed with the other endpoint's hostname or IP address, which will be looked up to verify the authenticity of its public key. The TLS client usually knows the server hostname, and when connecting to the server, it looks up the correct public key by the server hostname. In some constrained applications, such as sensor networks, the client may be configured with the server's IP address, in which case it looks up the correct public key by the server's IP address. Either way, this is quite natural because the TLS client looks up the correct public key by the same identifier that it uses for connecting to the server. 

On the other hand, it is less clear how the server in mutual RPK authentication can verify the client's public key. The server does not know the hostname of the connecting client, and the client's IP address is rarely a reliable identifier because most clients have dynamic IP addresses or are behind a NAT. Unlike X.509 client certificates, which contain the client hostname, the RPK is just a key and does not give any indication of the client identifier. In some situations, such as closed sensor networks, the client's IP address may be a sufficient identifier for the lookup. We will assume this method in our analysis.

\paragraph{Server authentication with DANE} The TLS RPK specification includes another method for TLS clients to verify the server's RPK: DNS-Based Authentication of Named Entities (DANE)~\cite{rfc7671,rfc6698}. DANE specifies a new DNS resource record (RR) type, TLSA, which binds a domain name to a public key or certificate. DANE can function as a replacement for the web PKI if, instead of trusting certificates issued by a CA, the client trusts the signed TLSA records in DNS Security Extensions (DNSSEC). When used to validate an RPK in TLS, a TLSA resource record will hold a public key or its hash. 

It is worth noting that DNSSEC does not require proof of possession of the corresponding private key when the public key is registered in a TLSA RR. Authorization of the domain owner is sufficient for the updates. DANE also allows many-to-many relations between the domains and public keys. A single hostname can have multiple TLSA RR entries with different public keys, and different domains can have TLSA RR entries with the same public key.

\paragraph{Client authentication with DANE} DANE currently does not support client authentication. However, there is an ongoing effort to standardize a mechanism for TLS servers to verify the client RPK or certificate with DANE~\cite{ietf-dance-client-auth-05}. The clients are identified by domain names, and the proposal defines a new extension to the TLS protocol to convey the TLS client identity to the server~\cite{ietf-dance-tls-clientid-03}.

The proposed extension serves multiple purposes: \begin{enumerate*}[label=(\roman*)] \item TLS client includes an empty DANE client identity extension in the \textit{ClientHello} to indicate that it has an identity that can be authenticated with DANE. \item TLS server includes an empty DANE client identity extension in the \textit{CertificateRequest} message to indicate that it is willing to authenticate the client's identity with DANE. \item TLS client includes a non-empty DANE client identity extension in its \textit{Certificate} message. The non-empty extension contains the encrypted client domain name, which the server should then use to verify the authenticity of the client's RPK with DANE. \end{enumerate*} While the above description is focused on TLS 1.3, which introduced the \textit{EncryptedExtensions}, the proposal also considers TLS 1.2, where the client identity would be communicated unencrypted in the \textit{ClientHello} and \textit{ServerHello} messages.

\subsection{Analysis of TLS RPK Implementations}

Many open-source TLS libraries (for example, OpenSSL, wolfSSL, and GnuTLS) support TLS RPK. Their approaches to handling the raw public keys differ somewhat from each other. OpenSSL~\cite{openssl} allows an application to invoke the \lstinline{SSL_get0_peer_rpk} function to learn and validate the received raw public key. WolfSSL~\cite{wolfssl} allows applications to register a callback function that validates the received raw public key. GnuTLS~\cite{gnutls}, on the other hand, maintains a database of trusted RPKs, which can be added with the \lstinline{gnutls_store_pubkey} function. It also supports trust on first use (TOFU), where the endpoints store the received RPKs and verify them on subsequent connections with the same endpoint. (The verification can be explicitly invoked by calling the \lstinline{gnutls_verify_stored_pubkey} function.)

Libcoap~\cite{libcoap}, a library for the Constrained Application Protocol (CoAP), supports RPK authentication of client Internet-of-Things (IoT) devices. However, the library documentation does not explain how a server can distinguish between client identities. When a client IoT device presents an RPK instead of an X.509 certificate, libcoap passes the RPK and the fixed CN string "RPK" to a callback function (registered with \lstinline{coap_dtls_cn_callback_t}). It is up to the application to determine the client's identity from the received public key. 

Some of these libraries also support DANE for authentication. OpenSSL~\cite{openssl} allows clients to input server TLSA RRs with the \lstinline{SSL_add_expected_rpk} function. Gnu\-TLS~\cite{gnutls} allows the clients to invoke the \lstinline{dane_verify_crt} function to verify the received server RPK via DANE. Apart from this DANE support, the libraries do not provide guidance to application developers on how to confirm that the received RPK belongs to the correct entity. The developers are left to implement their custom validation methods in the callback functions.


\section{Formal Model of TLS RPK}
\label{model}

This section explains the formal modeling methods used in the paper, describes the  model of TLS RPK, including the TLS handshake and the methods for binding the endpoint identities to their public keys, and specifies the main security goals for the verification.

\subsection{Symbolic Modeling Methods}

ProVerif~\cite{blanchet2018proverif} is a symbolic modeling tool used to verify security protocols. It takes a protocol model written in applied pi calculus and considers all possible execution paths in a protocol to find potential attacks. ProVerif uses the Dolev-Yao attacker model \cite{dolev1983security}, where the attacker controls the network and can read, modify, and inject messages. Security properties of a protocol are formalized as queries that are evaluated against the protocol model. Queries in ProVerif are written in terms of events. In particular, authentication is formalized as a correspondence between local events in the different concurrent processes. Proverif models are symbolic, and they treat cryptographic primitives, such as signatures and encryption, as abstract functions that are assumed secure. Unlike computational models~\cite{blanchet2007computationally}, the symbolic approach does not consider the probability of breaking cryptographic primitives. The symbolic approach can find logical flaws in the protocol design.

TLS has previously been analyzed with symbolic modeling tools. Bhargavan et al.~\cite{bhargavan17} used ProVerif to model TLS 1.3 draft 18, as well as the composition of TLS versions 1.3 and 1.2, to study the security properties in scenarios where clients and servers support both versions for backward compatibility. Cremers et al.~\cite{cremers17} modeled TLS 1.3 draft 21 with the Tamarin~\cite{meier2013tamarin} symbolic modeling tool. They analyzed, among other things, the key exchange mechanisms and session resumption with pre-shared keys. While earlier attempts at symbolic formal analysis of TLS have analyzed many protocol variants and found attacks, none explicitly study TLS with RPKs. The current paper aims to study the security properties of TLS when raw public keys are used for authentication.

\subsection{Model Overview}

We modeled server authentication and mutual authentication in TLS RPK with several key validation mechanisms. Since TLS RPK \textit{Certificate} messages do not contain full certificates issued by a CA, an out-of-band mechanism is needed for binding an endpoint’s identity to its public key. The TLS RPK standard discusses two out-of-band mechanisms to achieve this binding: pre-configured keys and DANE. We modeled the following TLS RPK scenarios:

\begin{enumerate}
    \item Server authentication with DANE in TLS RPK
    \item Mutual authentication with DANE in TLS RPK
    \item Mutual authentication with pre-configured keys in TLS RPK
\end{enumerate}

The ProVerif models are available for review online\footnote{\url{https://github.com/Mariam-Dessouki/tls-rpk}}.

\subsection{TLS Handshake Model}

We wrote our own model of the TLS handshake with a focus on the features relevant to TLS RPK. In other respects, the handshake model is similar to previously published formal models of TLS. It closely follows the message flows, contents, and key derivation in the TLS 1.3 standard. There are two main processes: the server and client process. After the \textit{ClientHello} and \textit{ServerHello} messages are sent, both endpoints derive the master secret. All the following messages are encrypted and authenticated with keys created from the master secret. The server sends the client its raw public key in the \textit{Certificate} message, a signed transcript hash in the \textit{CertificateVerify} message, and a MAC of the transcript hash in the \textit{Finished} message. The signature in the \textit{Certificate\-Verify} message proves to the client that the server possesses the private key corresponding to the RPK. 

\begin{figure}[t]
    \begin{subfigure}[t]{0.40\textwidth}
        \centering
        \vspace{0pt}
        \includegraphics[scale=0.6]{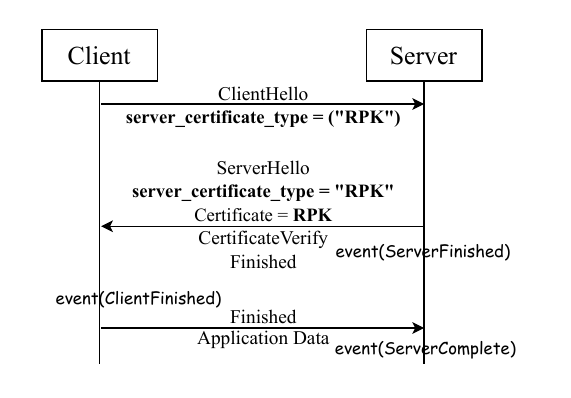}%
        \vspace{38pt}
        \caption{Events in the TLS RPK models}
        \label{fig:tls1.3-server-rpk-events}
    \end{subfigure}%
    \begin{subfigure}[t]{0.60\textwidth}
        \centering
        \vspace{0pt}
        \includegraphics[scale=0.6]{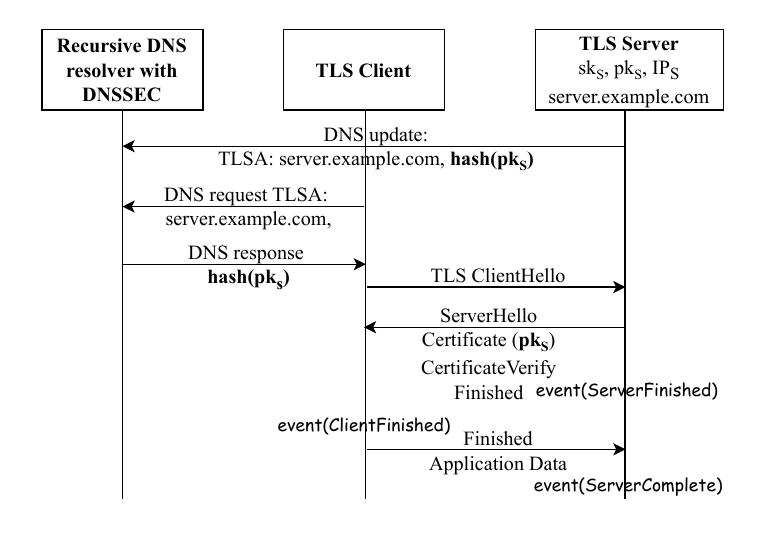}
        \caption{DANE registration}
        \label{fig:tls1.3-mutual-rpk-dane}
    \end{subfigure}
    \caption{Model details for TLS 1.3 authentication with RPK}
    \label{fig:tls1.3-rpk-model}
\end{figure}

Figure~\ref{fig:tls1.3-server-rpk-events} shows the client and server events that we will use to formulate the security goals. The server and client emit the \lstinline|ServerFinished| and \lstinline|ClientFinished| events when sending their respective \textit{Finished} messages. \lstinline|ClientFinished| signals that the client has received the final handshake message and accepted the server authentication. 

In the models with mutual authentication, the client sends its own \textit{Certificate} and \textit{CertificateVerify} messages before its \textit{Finished} message. In this case, the additional \lstinline|ServerComplete| event indicates that the server has accepted the client authentication.

\subsection{RPK Binding Model}

In addition to the TLS handshake, the TLS RPK authentication requires an out-of-band mechanism for binding the server's raw public key to its identity, such as its domain name. This part of our model is novel compared to the previously published TLS models. 

\paragraph{DNS Registration in DANE}

The DNS process receives queries and sends back signed resource records in the response. Additionally, the DNS process receives updates to the names. Each TLS endpoint can create TLSA resource records that bind its domain name to its public keys. The DNS update requires authorization, which can take many forms. For example, the domain owner could log in on the domain registrar’s web portal to edit its domain’s DNS records. We model the authorization in a relatively abstract way as a table of domain names and corresponding credentials. When an endpoint wants to update the TLSA resource records for its domain name, it sends an update message to the DNS protected by the credential.

A domain in the model can have multiple TLSA resource records with different public keys, and multiple domains can have the same public key in their resource records. DNS allows such many-to-many relationships between domains and public keys. 

When a client wants to connect to a server, it initially knows the server name. The client queries the DNS for the TLSA records associated with the server name. Upon receiving the server RPK in the TLS handshake, the client checks that it matches one of the values in the TLSA records. The DNS update and request are shown in Figure~\ref{fig:tls1.3-mutual-rpk-dane}. 

We also model mutual authentication with the proposed client name extension (see Section~\ref{sec:peer-rpk}). The server receives the client name, in addition to the RPK, in the Certificate message. The server queries the DNS for the TLSA records associated with the received name and checks that the received RPK matches one of them. 

Compromised or attacker-owned domains are modeled by leaking the credentials of some endpoints to the attacker. The event \lstinline|CompromiseDomain(d)| indicates that the domain's update credential has been given to the attacker. While the compromised domain itself cannot expect much security from DNS or DANE, the existence of some compromised domains should not endanger the security of other domains.

\paragraph{Pre-Configured Keys} 

In the pre-configured keys scenario, we model mutual authentication where each endpoint has a table of bindings between endpoint identifiers and their public keys. For a concrete example, consider IoT hubs and devices identified by IP addresses. The device IP address and public key are registered to the hub, and the device is pre-configured with the hub IP address and public key. 

An IoT device typically connects to one hub while the hub receives connections from multiple clients. In the pull communication pattern, however, the IoT hub could initiate connections to the devices. To cover all communication patterns, we model the relation between the TLS servers and clients as many-to-many. Thus, each TLS RPK server endpoint may be pre-configured with the public keys of many clients, and each client may be pre-configured with the public keys of many servers. 

We also model the pre-registration step and allow the attacker to register its own devices to the hub. This extension to the model was inspired by the DANE model discussed above and turned out to be essential for the security analysis.

\subsection{Security Goals}

The main security goal of the TLS handshake is authentication, which is expressed as a correspondence between the events in the client and server processes.

The client should only accept the TLS connection if it is established with the server to which the client intended to connect. The following security query represents server authentication in ProVerif. 

\begin{lstlisting}[language=ProVerif,numbers=none,captionpos=t,caption={Server authentication goal},label={lst:server-auth}]
    query s_domain:Id_t, rpk:PK_t, ms:K_t;
    event(ClientFinished(s_domain, rpk, ms)) 
    ==> inj-event(ServerFinished(s_domain, rpk, ms))
        || event(CompromiseDomain(s_domain)).
\end{lstlisting}

The query states that if the (possibly anonymous) client accepts a TLS connection to a server with the server name \lstinline{s_domain}, server public key \lstinline{rpk} and master secret \lstinline{ms} for protecting the TLS connection, then the server with the name \lstinline{s_domain} and public key \lstinline{rpk} must have derived the same master secret. An exception is allowed for the case where the domain is compromised, i.e., the attacker has compromised the DNS update credentials, or it is an attacker-owned domain to begin with.

A second query represents client authentication in the mutual authentication scenario.
 
 \begin{lstlisting}[language=ProVerif,numbers=none,captionpos=t,
    caption={Client authentication goal},label={lst:client-auth}]
    query s_domain,c_domain:Id_t, spk,cpk:PK_t, ms:K_t;
    event(ServerComplete(s_domain,c_domain,spk,cpk,ms))
    ==> event(ClientFinished(s_domain,c_domain,spk,cpk,ms))
        || event(CompromiseDomain(s_domain))
        || event(CompromiseDomain(c_domain)).
\end{lstlisting}

The query states that if a server with a name \lstinline|s_domain| and public key \lstinline|spk| accepts a TLS connection from a client with the client name \lstinline|c_domain|, client public key \lstinline|cpk| and master secret \lstinline|ms|, then the client should have already accepted the corresponding connection to the server. An exception is allowed for the cases where the server's or the client's DNS domain is compromised. 

Aditionally, there are secrecy goals related to the master secret. We omit them here for brevity and because the interesting results come from the queries above.


\section{Misbinding Attack in RPK Registration}
\label{misbinding}
\label{sec:misbinding_attacks}

This section presents the main results of our formal modeling and analysis. They are misbinding attacks that exploit weaknesses in how the standard binds endpoint identity to its public key.

\paragraph{Server Misbinding Against TLS RPK with DANE}

\begin{figure}[ht]
\includegraphics[width=1\textwidth]{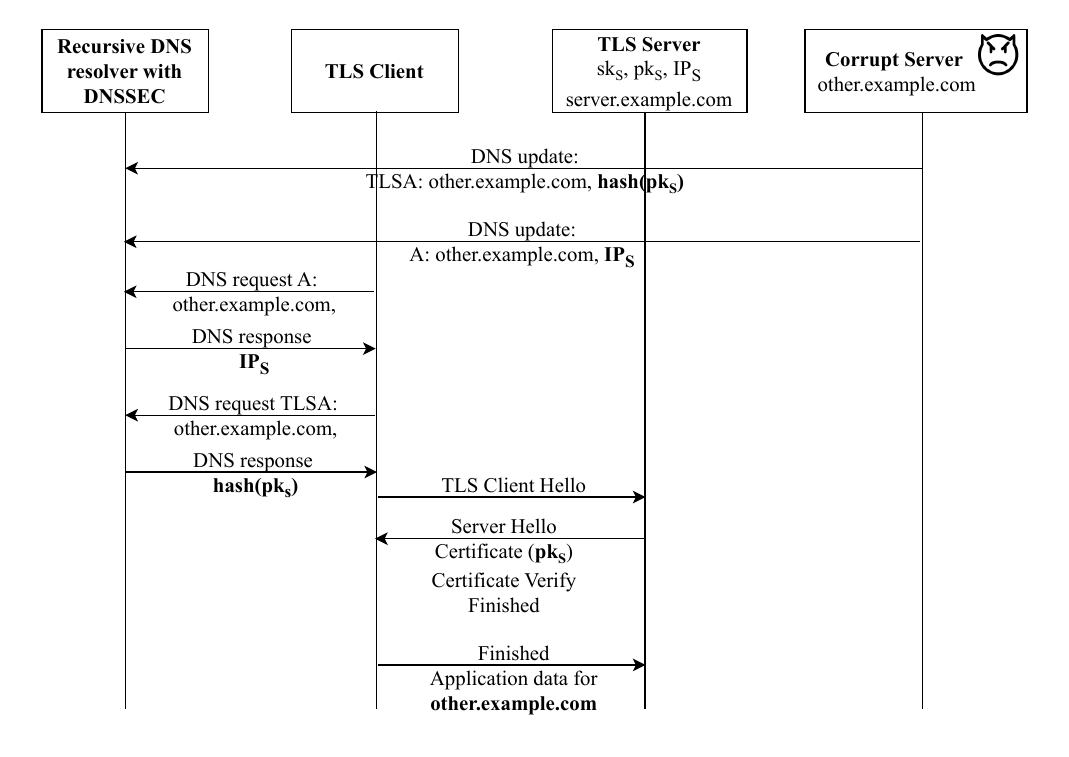}
\caption{Misbinding attack against TLS RPK with DANE, exploiting DNS registration}
\label{fig:attack1} 
\end{figure}

The query for server authentication in Listing~\ref{lst:server-auth} fails when DANE is used for authentication. The attack is illustrated in Figure~\ref{fig:attack1}. A malicious domain owner, \lstinline|other.example.org| in the figure, has full control over its own DNS zone and TLSA resource records. It can thus write the hash of the public key of the honest server \lstinline|server.example.com| into a TLSA resource record of its own domain name. A client that wants to communicate with the malicious domain receives the public key belonging to the honest server. The attacker then redirects inbound connections from the malicious domain to the honest server. The client connects to the wrong server but successfully verifies the signature in the \textit{CertificateVerify} message because the received RPK matches the one in the TLSA resource records of \lstinline|other.example.org|. The client accepts the connection with the unintended server, and hence the server authentication goal fails. This attack is known as a misbinding~\cite{sigma,sethi19} or unknown key share attack~\cite{uks}.

The attack is easier to implement than one might think. First, the DNS registrar and server are not compromised. Instead, the malicious domain owner controls the DNS records for a domain which it has legitimately registered. Second, no advanced network attacks, such as forwarding connections or manipulating the network routing, are necessary. Instead, the malicious domain owner can redirect the connections to the honest server by copying the IP address of the honest server to an A resource record of \lstinline|other.example.org|. This will cause the client to connect to the honest server instead of the intended one. Thus, no special skills or access to a specific network segment are needed for the attack implementation. 

The attack clearly violates the security goals. The practical consequences of the misbinding attack can be difficult to understand because the client is willingly connecting to the malicious domain. This kind of behavior is, however, quite common on the Internet. For example, web browsers and email relays connect to untrusted servers, which could behave maliciously. We will present a more detailed scenario related to mail servers when discussing the attack implementation in Section~\ref{sec:implementation}.

\paragraph{Server Misbinding with Pre-Configuration}

The query in Listing~\ref{lst:server-auth} also reveals an attack in the case of pre-configured public keys. We explain the attack in the context of an example IoT application. Consider a network of IoT devices, identified by their IP addresses or domain names, and a hub to which the devices are registered. Following the pull communication pattern, the IoT hub connects to the devices to collect data from them. Let \lstinline|device1| be an honest IoT device. A malicious user can copy the public key of \lstinline|device1| and register a new (real or imaginary) device \lstinline|device2| to the same hub with the same public key. When the hub polls \lstinline|device2| for data over TLS RPK, the network attacker rewrites the device IP address in the network layer and redirects the connection to \lstinline|device1|. The RPK received from \lstinline|device1| in the TLS handshake matches the key registered for \lstinline|device2|. Thus, the IoT hub falsely accepts the connection as coming from \lstinline|device2|. The IoT hub then records the received data as coming from the wrong device.

\paragraph{Server Misbinding for Multi-Named Server}

\begin{figure}[ht]
\includegraphics[width=1\textwidth]{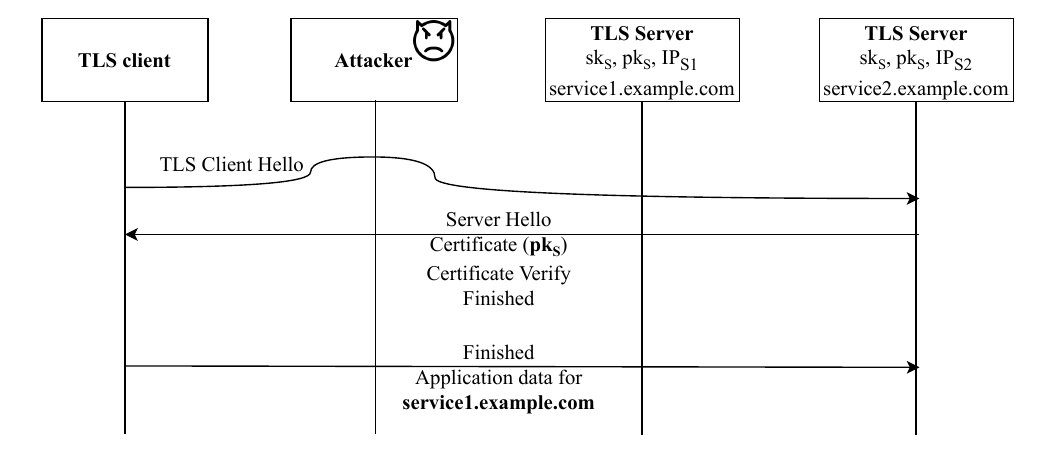}
\caption{Misbinding attack against TLS RPK, exploiting multi-name server}
\label{fig:attack2} 
\end{figure}

As a matter of fact, there is a simpler attack on server authentication where no malicious registration is needed. DNS allows for a many-to-many relation between domains and public keys. Suppose there are two services: \lstinline{coaps://service1.example.com} and \lstinline{coaps://service2} \lstinline{.example.com} that share the same public key as shown in Figure~\ref{fig:attack2}. 
When the honest device connects to the server, a network attacker can redirect the connection from service1 to service2, or vice versa. Since the public keys are identical, the client accepts the RPK received in the TLS handshake even when connected to the wrong server.  

There are practical scenarios where two servers could have the same public key. Hosting multiple online services on the same physical server is common, and cloud application gateways can hide any number of services behind the same application gateway and public IP address. The two different services could even have the same backend authentication server and the same client accounts for authenticating the user inside the TLS tunnel. The attack results in the client sending the application-layer data to the wrong service.

This attack is similar to one of the attacks presented in ALPACA~\cite{alpaca21}. If a certificate with multiple hostnames or a wildcard pattern in the common name field is used in communication, an attacker can reroute the connection to an unintended server with an SNI that matches one of the common name values in the certificate. The certificate would still be valid, but the client would be communicating with the wrong server.

\paragraph{Weakness in the TLS RPK specification}

The misbinding attacks presented in this section can be prevented by optional features in the TLS handshake. In TLS, the client can optionally send the \textit {server name indication} (SNI) in the \textit{server\_name} extension of the \textit{ClientHello} message. The server may optionally use the received SNI to select the public key, which implicitly means that the server checks the SNI value and rejects unrecognized names. If the client sends the extension and the server checks it, the misbinding attacks are prevented. The security queries succeed when both actions are added in the models. However, since these actions are optional in the standard, there is a lot of room for insecure implementations and usage.


\section{Analysis of Client Authentication}
\label{mutual}

So far, we have discussed misbinding of the server identity, causing the client to connect to the wrong server. What about misbinding of the client identity in mutual authentication? The misbinding attacks in the literature typically have two possible directions, and the same could happen in TLS RPK. This section analyzes whether there are similar weaknesses in client authentication.

\paragraph{No Client Misbinding Against TLS RPK with DANE}

Client authentication using DANE was modeled based on the Client Authentication proposal~\cite{ietf-dance-client-auth-05}, which we discussed in Section~\ref{sec:peer-rpk}. When RPK is used for client authentication, the client must send the \textit{ClientName} extension \cite{ietf-dance-tls-clientid-03} in the \textit{Certificate} message. Knowing the client domain name allows the server to look up the client's public key from TLSA resource records in the DNS, and this provides a secure binding between the client domain name and its public key. Thus, the client authentication with DANE is not susceptible to a misbinding attack. The critical difference between client and server authentication with DANE is that \textit{ClientName} is mandatory while \textit{SNI} is optional and, even when the client sends the \textit{SNI}, the server may ignore it.

\paragraph{Client Misbinding with Pre-Configuration}

An attack was found on client authentication with pre-configured public keys. Consider again the IoT hub and devices. In this case, the devices send push messages to the IoT hub, reversing the direction of the TLS connections compared to our earlier scenario. The hub identifies the devices by their IP address, which is visible in the IP header but not protected by the TLS handshake. A malicious user can register a new (real or imaginary) IoT device with the public key of another, honest device. When the honest device connects to the server, the network attacker performs address translation on the client's IP address, replacing the honest device's IP address with the malicious user's one. The IoT hub checks that the RPK received in the TLS handshake matches the one registered for the malicious user's device, and since it does, the hub accepts the connection. Consequently, the hub mistakenly believes that the data received over the TLS connection comes from the malicious user's device.


\section{Attack Implementation}
\label{sec:implementation}

We successfully verified the practicality of the server misbinding attacks discovered by the formal analysis across three popular TLS libraries: OpenSSL, GnuTLS, and wolfSSL. Since TLS RPK with DANE and server authentication has the most complete and well-specified key registration solution, we focused on implementations that support DANE.

Initially, we deployed both an honest server and an attacker-controlled server, where the IP address and TLSA record of the attacker-controlled server were modified to match those of the honest server. The attack proved to be relatively straightforward. In practice, the attacker-controlled server was not even used. It was sufficient to modify the DNS entries of the compromised server. The unsuspecting client was redirected to the honest server and could not detect the misbinding. 

Subsequently, we discovered a rare public mail server that supports TLS RPK for server authentication and has its TLSA record published in DNS. We set up a test SMTP client and server that use StartTLS with RPK and DANE. We then copied the public mail server's IP address and TLSA resource record to the DNS records of our test server. This caused the unsuspecting test client to connect to the public mail server instead of the intended test server. 

The practicality of such attacks against TLS RPK in real-world applications depends on the underlying TLS libraries and on how the applications interact with them. For example, in OpenSSL, the client application can call the \textit{SSL\_set\_tlsext\_host\_name} function to include an SNI into the \textit{ClientHello} message. It is then up to the server implementation to verify or ignore the SNI value. The public mail server in our experiment accepted any SNI value without raising an alert. Even servers that adhere strictly to the standard specification~\cite{rfc8446} might only issue a warning-level TLS alert upon encountering an unrecognized SNI but nevertheless continue with the connection.

The outcome of the attack also depends on the application layer. For example, HTTPS clients (HTTP version 1.1 and above) send a Host header to the server, and the server may reject requests for an unrecognized server name in the header. Some reverse proxies and TLS gateways, however, are configured with a default forwarding rule that ignores the Host header.


\section{Solutions for Preventing Misbinding}
\label{solution}

To prevent the misbinding attack, the endpoints of the session must agree on the authenticated identifiers associated with the session. In other words, the endpoint identities should be bound to the TLS session \cite{uks,sigma}. Several solutions can be employed for this purpose. 

\subsection{Sending the Identities in the TLS Handshake}

As already mentioned, one way to bind the server or client identity to the session is to send them in the handshake messages. The SNI in the \textit{ClientHello} message exists for this purpose. The SNI will automatically become part of the transcript hash, which is authenticated in the handshake. To prevent misbinding, all that is needed is to make the \textit{server\_name} extension mandatory in TLS RPK and require the server to check the received value. 

A disadvantage of this solution is that it is not easy to transition to a stricter standard in an open system such as the Internet. The client cannot ensure that the server checks the received SNI, and if the server follows the new stricter rules, compatibility with legacy clients might break.

The equivalent extension for the client identity is currently a proposal and not yet a part of any standard. The DANE Client Identity draft~\cite{ietf-dance-tls-clientid-03} would make it mandatory for the client to send the \lstinline{dane_clientid} extension when using an RPK. Since DANE for client authentication is still in the early draft stage, it would be possible to establish stricter rules for it. A disadvantage of the proposal is that it is specific to DANE and does not solve the problem in key pre-configuration and other out-of-band mechanisms.

Another possibility is to add new server and client identity extensions to the TLS specification for TLS RPK. They could be similar to the extensions \lstinline{server_certificate_type} and \lstinline{client_certificate_type} that were added to support TLS RPK. Sending the extension should probably be mandatory. Most importantly, checking the value should be mandatory at the receiving endpoint, raising a fatal alert if the value is not recognized.  

The privacy of the client and server needs to be considered. The server identity would be sent as plaintext in the \textit{ClientHello} message. The privacy of the client identity would depend on the TLS version. In TLS 1.3, the client could send its identity as an encrypted extension of the \textit{Certificate} while, in TLS 1.2, it would have to be sent in plaintext.

 \subsection{Self-Signed Certificates}

One reason why the SNI is optional in TLS~\cite{rfc8446} is that, in certificate-based authentication, the server certificate identifies the server. The endpoints agree on the server name in the certificate, which is sufficient to prevent server misbinding. 

TLS RPK could get the same benefit from self-signed certificates. DANE already supports storing a certificate hash in the TLSA RR. If an endpoint sends a self-signed certificate with the correct domain name (or another correct identifier) in the \textit{Certificate} messages and the other endpoint validates the certificate following the PKIX validation rules \cite{rfc5280}, including checking the subject name, then misbinding of the subject is prevented. DANE already supports this solution with the Certificate Usage PKIX-EE(1). Self-signed certificates would also work well together with pre-configured public keys. 
 
While self-signed certificates prevent the misbinding attack, they still have the same problems that TLS RPK was set out to fix. The entire certificate needs to be transmitted, and the endpoints must have full ASN.1 parsers for the X.509 standard.

\subsection{Verify the Identities Inside the TLS Tunnel}

Yet another solution is to verify the endpoint identities in the application layer inside the established TLS tunnel. As mentioned earlier, the Host header in the HTTP request can prevent misbinding if the server carefully checks the value. While such application-layer mechanisms can intentionally or accidentally fix the security issue, the solution does not work universally for all applications of TLS RPK. It also seems that the architecturally more correct approach is to fix TLS RPK security flaws in the TLS RPK protocol.


\section{Conclusion}
\label{conclusion}
We have modeled and analyzed the TLS Raw Public Key (RPK) mode, focusing on scenarios where identity of the server and client are linked to their public keys through various out-of-band mechanisms, including secure DNS. We performed verification of the protocol, which led to the discovery of identity misbinding attacks. In addition to presenting the verification results and misbinding vulnerabilities, we show their significance by presenting concrete attack scenarios and providing guidance for enhancing the robustness of TLS RPK standard and implementations.

%
%
%
\bibliographystyle{splncs04}
\bibliography{citations}
\end{document}